\documentstyle[12pt,epsfig]{article}
\title{QUARK-ANTIQUARK BETHE-SALPETER FORMALISM, SPECTRUM AND REGGE 
TRAJECTORIES}
\author{\large G. M. Prosperi \\[3mm]
\em Dipartimento di Fisica, Universit\`a di Milano,  \\
\em INFN, sezione di Milano,\\
\em via Celoria 16, I20133 Milano, Italy}
\date{Sept 4, 1998}

\begin{document}
\maketitle

\begin{abstract}
    Starting from a path integral representation of appropriate 
4-point and 2-point gauge invariant Green functions and from the "Modified 
Area Law" model 
, a $q \bar q$ 
Bethe-Salpeter like equation and a related Schwinger-Dyson equation can be 
obtained. From such equations an effective relativistic Hamiltonian can 
be derived by standard methods and then applied to the determination of the 
meson spectrum . The entire known heavy-heavy and heavy-light spectra 
and the lowest light-light 
Regge trajectories are rather well reproduced in terms of four parameters
alone, the light quark masses being fixed a priori on typical current values.  
\end{abstract}


\section{Introduction}

   In this paper I want to revue a Bethe Salpeter formalism 
in QCD, which has been have developed recently in Milano. Such 
formalism, if not completely derived from first principles, rests, however, 
only on some non controversial assumptions on the Wilson loop correlator 
\cite{prosp96}. It generalizes a method introduced previously for the case of 
the heavy quark potential \cite{prspcomo1} and takes advantage of 
appropriate Feynmann-Schwinger like path integral representations for the 
QCD Green functions. 
Its three dimensional reduction has been recently applied to the light-light 
and heavy-light quark-antiquark spectrum with very encouraging results 
\cite{mio}. Beside the heavy quarkonia, the lowest Regge trajectories for 
the triplet $u\bar u, \ u\bar s, \ s\bar s$ states are very well reproduced 
in slope and intercepts and the known spin averaged light-heavy spectrum is 
obtained up to a mean deviation of about 10 MeV. Notice that it is very 
difficult to obtain a similar result in the frame of a single model. In 
fact in our case the strong coupling constant $\alpha_{\rm s}$, the string 
tension $\sigma$ and the heavy quark masses $m_b$ and 
$m_c$  were already completely determined by 
the potential fits (apart the possibility of a very small rearrangement), 
while the light quark masses have been fixed  a priori on typical 
current values ($m_u=m_d= 10 \, {\rm MeV}, \  m_s=200 \, {\rm 
MeV}$). In particular the heavy-light sector is completely parameter 
free.

    The basic objects from which we start are 
the ordinary gauge invariant 
4-point and 2-point Green functions $G^{\rm gi}(x_1,x_2,y_1,y_2)$ and
$G^{\rm gi}(x-y)$, to which we refer as the ``first order''  
functions. To these first order functions certain ``second order'' ones 
$H^{\rm gi}(x_1,x_2,y_1,y_2)$ and $H^{\rm gi}(x-y)$ can be related.
It is for the second order functions that the mentioned path integral 
representations can be constructed.

  The important aspect of the above representations is that the 
gauge field occurs simply trough Wilson correlators 
 \begin{equation}
W[\Gamma] = {1\over 3}
   \langle {\rm Tr} {\rm P} \exp \{ i g  \oint_\Gamma dx^{\mu} 
   A_{\mu} \}  \rangle  \ .
\label{eq:loop}
  \end{equation}
associated to loops $\Gamma$ made by quark or antiquark world lines and 
``Schwinger strings''. In principle these correlators should 
determine the whole dynamics. Unfortunately, due to confinement and the 
consequent failure of a purely perturbative approach, a consistent analytic 
evaluation of $W$ from the Lagrangian alone is not possible today. However 
combining incomplete theoretical arguments and lattice simulation information 
various reasonable models can be attempted. 

  The most naive but at the same time less arbitrary assumption consists
in writing $i\ln W$ as the sum of its perturbative expression and an 
area term (modified area law (MAL) model)
  \begin{equation}
i \ln W = i (\ln W)_{\rm pert} + \sigma S_{\rm min}   \,  ,
\label{eq:wl}
   \end{equation}
where the first quantity is supposed to give correctly the short range 
limit the second the long range one. 

  Eq. (\ref{eq:wl}) is our starting point. In principle any more 
sophisticated model could be used in the context, at the condition that it
preserves 
certain general properties of functional derivability of the definition 
(\ref{eq:loop}). In practice not even (\ref{eq:wl}) can be treated exactly. 
Actually we shall replace the minimal surface $S_{\rm min}$ by what can be
called its ``equal time straight  
line approximation'', which we shall  explain later.
 
  In connection with (\ref{eq:wl}) it is convenient to consider a 
third type of Green functions $H(x_1,x_2,y_1,y_2)$ and $H(x-y)$ which are 
obtained from $H^{\rm gi}(x_1,x_2,y_1,y_2)$ and $H^{\rm gi}(x-y)$ 
by omitting in their path integral representation contributions 
related to the Schwinger strings. In the limit of vanishing $x_1-x_2,\ 
y_1-y_2$ or $x-y$ such new quantities coincide with the original ones and
are completely equivalent for what concerns the determination of  bound states, condensates, 
chiral symmetry breaking, etc.
  
  For $H(x_1,x_2,y_1,y_2)$ and $H(x-y)$, an inhomogeneous Bethe-Salpeter 
equation and a Dyson-Schwinger equation, respectively, can be derived in 
the configuration space, with kernel obtained as an 
expansion in $\alpha_{\rm s}$ and $\sigma$. 
Such equations can be rewritten in a more conventional form in the momentum
space  and as such applied e. g. to the spectrum of the mesons. To this aim
in principle one should solve the DS equation first and use the resulting
propagator in the BS equation. In practice a direct treatment of their full four dimensional 
expressions seems to be a very difficult task. However, if one neglects the
pseudo-scalar mesons, a three dimensional reduction of the 
BS-equation seems to be appropriate. This can be obtained by standard
methods in the form of an eigenvalue equation for an effective squared mass 
operator or simply a relativistic Hamiltonian  \cite{prosp96,prspcomo1}.   

   For the case of the pseudo-scalar mesons a complete four-dimensional treatment would be 
imperative. In fact the use of the free quark propagator implied in the
three dimensional reduction can not be suitable in this case, for the strict interplay 
existing between zero mass Bethe-Salpeter wave function and the propagator in the chiral 
limit; we shall limit to few comments.

   It should be mentioned that our formalism is strictly related from 
different point of view to to the works of ref. \cite{tubolsson}.
  
   The plan of the paper is the following one. We discuss the gauge invariant 
Green functions and their path integral representations in sect. 2; modified 
area law model and straight line approximation in sect. 3; Bethe-Salpeter, 
Dyson-Schwinger equations and the problem of chiral symmetry in sect. 4. 
In sect. 5 we discuss the three dimensional reduction of the 
BS-equation and its application to the determination of the spectrum and 
the Regge trajectories.


\section{Green functions and Feynman-Schwinger representations}

     The quark-antiquark and the single quark gauge invariant Green 
functions are defined as 
\begin{eqnarray}
G^{\rm gi}(x_1,x_2,y_1,y_2) &=&
\frac{1}{3}\langle0|{\rm T}\psi_2^c(x_2)U(x_2,x_1)\psi_1(x_1)
\overline{\psi}_1(y_1)U(y_1,y_2)  \overline{\psi}_2^c(y_2)
|0\rangle =
\nonumber\\
&=& \frac{1}{3} {\rm Tr_C} \langle U(x_2,x_1)
 S_1(x_1,y_1;A) U(y_1,y_2)
\tilde S_2(y_2,x_2;-\tilde A) \rangle
\label{eq:qqgauginv}
\end{eqnarray}
(plus an annihilation term in the equal flavor case) and 
\begin{equation}
G^{\rm gi}(x-y) =
\langle0|{\rm T}U(y,x)\psi(x)
\overline{\psi}(y)
|0\rangle = i {\rm Tr_C} \langle U(y,x)
 S(x,y;A) \rangle \, ,
\label{eq:qgauginv}
\end{equation}
where $\psi^c$ denotes the charge-conjugate fields, the tilde and 
${\rm Tr_C}$ the transposition and the trace respectively over the 
color indices alone and $U$ the path-ordered gauge string (Schwinger 
string)
\begin{equation}
U(b,a)= {\rm P}  \exp \left \{ig\int_a^b dx^{\mu} \, A_{\mu}(x) 
\right \}\, .
\label{eq:col}
\end{equation}
Notice that the integration in (\ref{eq:col}) is along an arbitrary line joining $a$ to 
$b$ (which usually we shall not specify explicitly), $S$, $S_1$ and $S_2$ are 
the quark propagators in the external gauge field $A^{\mu}$ and the angle 
brackets denote average on the gauge variable alone (weighted in principle
with the determinant  $M_f(A)$ resulting from the explicit integration of 
the fermionic fields). 

    The  propagator  $S$ is supposed to be defined by the equation (we shall 
suppress indices specifying the quarks, as a rule, when dealing with single 
quark quantities)
\begin{equation}
( i\gamma^\mu D_\mu -m) S(x,y;A) =\delta^4(x-y)
\label{eq:propdir}
\end{equation}
\noindent
and the appropriate  boundary conditions. This can be rewritten as 
\cite{prosp96}
\begin{equation} 
S(x,y;A) = (i \gamma^\nu D_\nu + m) \Delta^\sigma(x,y;A) ,
\label{eq:fsord}
\end{equation}
in terms of a ``second order'' propagator defined in turn by the equation
\begin{equation}
(D_\mu D^\mu +m^2 -{1\over 2} g \, \sigma^{\mu \nu} F_{\mu \nu})
\Delta^\sigma (x,y;A) = -\delta^4(x-y) \, ,
\label{eq:propk}
\end{equation}
with $\sigma^{\mu \nu} = {i\over 2} [\gamma^\mu, \gamma^\nu]$.

    After replacing (\ref{eq:fsord}) in (\ref{eq:qqgauginv}) and
(\ref{eq:qgauginv}), using an appropriate derivative it is possible
to take the differential operator out of the angle brackets and write
\footnote{Given a functional $\Phi[\gamma_{ab}]$
 of the curve $\gamma_{ab}$ with ends $a$ and $b$, let us assume that the 
 variation of $\Phi$ consequent 
 to an infinitesimal  modification  of the curve $\gamma \rightarrow \gamma
 +\delta \gamma $ can be expressed as the sum of various terms
 proportional respectively to $\delta a$, to $\delta b$ and to the single
 elements $\delta S^{\rho \sigma}(x)$ of the surface swept by the curve.
 Then, the derivatives $ \bar{\partial}_{a^{\rho}}$,
 $ \bar{\partial}_{b^{\rho}}$ and $\delta / \delta 
 S^{\rho \sigma}(x)$ are defined by the 
 equation $ \delta \Phi = \delta a^{\rho}
 \bar{\partial}_{a^\rho} \Phi 
 + \delta b^{\rho} \bar{\partial}_{b^\rho} \Phi   + {1 \over 2}\int_{\gamma} 
 \delta S^{\rho \sigma}(x) \delta  \Phi / \delta S^{\rho \sigma}(x)$. 
 For a Schwinger string we have $ \delta U(b,a) = \delta b^\rho ig A_\rho (b)
 U(b,a) - \delta a^\rho U(b,a) ig A_\rho (a) + {ig \over 2} \int_a^b 
 \delta S^{\rho \sigma}
 (z) {\rm P}(-F_{\rho \sigma} (z) U(b,a) $ and so
 $\bar{\partial}_{a^{\rho}} U = -ig U A_{\rho}(a) $,
 $\bar{\partial}_{b^{\rho}} U = ig A_{\rho}(b) U $ and $ {\delta \over 
 \delta S^{\rho \sigma}(z)} U = {\rm P}[-ig F_{\rho \sigma} (z) U] $.}
\begin{equation}
G^{\rm gi}(x_1,x_2; y_1,y_2) =-(i \gamma_1^\mu \bar{\partial}_{1 \mu} 
 + m_1) ( i \gamma_2^\nu \bar{\partial}_{2 \nu} +m_2) 
H^{\rm gi}(x_1,x_2;y_1,y_2) \, ,
\label{eq:qqeqg}
\end{equation}
%
\begin{equation}
G^{\rm gi}(x - y) = (i \gamma_1^\mu \bar{\partial}_{\mu} 
 + m) H^{\rm gi}(x-y) \, ,
\label{eq:qeqg}
\end{equation}
\noindent
with
\begin{equation}
H^{\rm gi}(x_1,x_2;y_1,y_2) = -{1\over 3} {\rm Tr _C}
\langle U(x_2,x_1) \Delta_1^\sigma (x_1,y_1;A)
 U(y_1,y_2) \tilde{\Delta}_2^\sigma (x_2,y_2;-\tilde{A})\rangle \, ,
\label{eq:qqeqh}
\end{equation}
%
\begin{equation}
H^{\rm gi}(x-y) = i {\rm Tr _C}
\langle U(y,x) \Delta ^\sigma (x,y;A) \rangle .
\label{eq:qeqh}
\end{equation}

  For the second order propagator we have the Feynman-Schwinger 
representation
\begin{equation}
\Delta^\sigma (x,y; A)= -{i \over 2} \int_0^\infty ds \int_y^x
   {\cal D} z \exp [-i \int_0^s d\tau {1\over 2} (m^2 
   +\dot z ^2)] {\cal S}_0^s {\rm P} \exp[ ig \int_0^s d \tau
   \dot z ^\mu A_{\mu}(z)] \, , 
\label{eq:partbis}
\end{equation}
\noindent
with 
\begin{equation}
{\cal S}_0^s = {\rm T} \exp \Big [ -{1\over 4} \int_0^s d \tau 
 \sigma^{\mu \nu} {\delta \over \delta S^{\mu \nu}(z)}
\Big ]\, ,
\label{eq:defop}
\end{equation}
${\rm T}$ and ${\rm P}$ being the ordering prescriptions along the path 
acting on the spin and on the color matrices respectively and $ \delta 
S^{\mu \nu} = dz^\mu \delta z^\nu - dz^\nu \delta z^\mu $ (the functional 
derivative being defined through an arbitrary deformation, $ z \rightarrow 
z + \delta z $, of the line connecting $a$ to $b$, see footnote) \cite{prosp96}. Replacing 
(\ref{eq:partbis}) in (\ref{eq:qqeqh}) and (\ref{eq:qeqh}) we obtain
\begin{eqnarray}
& &H^{\bf gi}(x_1,x_2;y_1,y_2)  = ({1 \over 2})^2 \int_0^{\infty} d s_1
\int_0^{\infty} d s_2
 \int_{y_1}^{x_1} {\cal D} z_1\int_{y_2}^{x_2} {\cal D} z_2 
\nonumber \\
& & \qquad \qquad \exp \Big \{  -{i \over 2} 
 \int_0^{s_1} d\tau_1 (m_1^2 +\dot{z}_1^2) - {i\over 2} \int_0^{s_2}
d\tau_2 (m_2^2 +\dot{z}_2^2)\Big \} \nonumber \\
& &  \qquad \qquad {\cal S}_0^{s_1} {\cal S}_0^{s_2}
{1\over 3} \langle {\rm Tr} \, {\rm P}
 \exp  \big \{ ig \oint_{\Gamma_{\bar q q}} dz^\mu A_{\mu} 
(z)   \} 
\rangle    ,
\label{eq:hqq}
\end{eqnarray}
%
\begin{eqnarray}
H^{\bf gi}(x-y)  &=& {1 \over 2} \int_0^{\infty} d s
 \int_y^x {\cal D} z
 \exp \big \{  -{i \over 2} 
 \int_0^{s} d\tau (m^2 +\dot{z}^2) \nonumber \\
& &  {\cal S}_0^s
 \langle {\rm Tr}\, {\rm P}
 \exp  \big \{ ig \oint_{\Gamma _q} dz^\mu A_{\mu} 
(z)   \} 
\rangle    .
\label{eq:hq}
\end{eqnarray}
\noindent
Here, the loop $\Gamma_{\bar q q}$ occurring in the 4-points function is 
made by the quark world line $\gamma_1$, the antiquark world line 
$\gamma_2$ followed in the reverse direction, and the two strings 
$x_1 x_2$ and $y_2 y_1$; the loop $\Gamma _q$ occurring 
in the 2-points function
is made simply by the quark trajectory $\gamma$ connecting $y$ to $x$ 
and the string $yx$.


\section{Wilson loop correlators}

     We now apply Eq. (\ref{eq:wl}) to evaluate the Wilson correlators  
for $\Gamma_{\bar q q}$  and  $\Gamma _q$. 

     At the lowest order the perturbative term can be written for any 
loop $\Gamma$

\begin{equation}
i (\ln W[\Gamma])_{\rm pert} = -{2 \over 3} g^2 \oint dz^\mu \oint dz^{\nu \prime}
D_{\mu \nu}(z-z^\prime)
\label {eq:perta}
\end{equation}
\noindent
$ D_{\mu \nu}(z-z^\prime) $ being the free gauge propagator. If we neglect 
the contribution coming from propagators connecting a point on a world-line 
to a point on a string or two point on the strings, we can write for 
$\Gamma_{q\bar q}$

\begin{eqnarray}
i (\ln W[\Gamma_{\bar q q}])_{\rm pert}&=& {4\over 3} g^2 \int_0^{s_1} 
 d \tau_1
\int_0^{s_2} d\tau_2 D_{\mu \nu}(z_1-z_2)
 \dot{z}_1^{\mu} 
\dot{z}_2^{\nu}-   \nonumber \\
& & \qquad \quad -{4\over 3} g^2 \sum_{j=1}^2 \int_0^{s_j} d \tau_j
\int_0^{\tau _j} d\tau_j^{\prime}
 D_{\mu \nu}(z_j-z_j^{\prime}) \dot{z}_j^{\mu} 
\dot{z}_j^{\prime\nu}
\label{eq:pertb}
\end{eqnarray}

\noindent
and for $\Gamma_q$

\begin{equation}
i (\ln W[\Gamma_q])_{\rm pert} =  -{4\over 3} g^2 \int_0^{s} d \tau
   \int_0^{\tau } d\tau^{\prime}
   D_{\mu \nu}(z-z^{\prime}) \dot{z}^{\mu} 
   \dot{z}^{\prime \nu}
\label{eq:pertc}
\end{equation}
Let us further consider for the moment the case in which  $\Gamma_{q \bar 
q}$ lies on a plane. Then $S _{\rm min}$ coincides simply with the portion 
of plane delimited by $\Gamma_{q \bar q}$ and it can be written, in a four 
dimensional language, \cite{prosp96}
\begin{eqnarray}
S _{\rm min}&=& \int_0^{s_1}
   d\tau_1  \int_0^{s_2} d\tau_2 \delta(z_{10}-z_{20})
   \vert {\bf z}_1 -{\bf z}_2\vert
   \epsilon (\dot{z}_{10})
   \epsilon (\dot{z}_{20}) \nonumber \\
  & & \qquad \quad  \int_0^1 d\lambda
   \Big \{ {\dot{z}_{10}^{ 2}} {\dot{z}_{20}^{ 2}} -
   (\lambda \dot{\bf z}_{1{\rm T}} \dot{z}_{20}
   + (1-\lambda) \dot{\bf z}_{2 {\rm T}}
   \dot{z}_{10} )^2 \Big \}^{1\over 2}  - \nonumber \\
 & & - \sum _{j=1}^2 \int_0^{s_j}
   d\tau_j \int_0^{\tau_j} d\tau_j^{\prime} \delta(z_{j0}-z_{j0} ^{\prime})
  \vert {\bf z}_j -{\bf z}_j ^{\prime}\vert
   \epsilon (\dot{z}_{j0})
   \epsilon (\dot{z}_{j0}^{\prime} ) \nonumber \\
 & & \qquad \quad  \int_0^j d\lambda
   \Big \{ {\dot{z}_{j0}^{ 2}} {\dot{z}_{j0}^{\prime 2}} -
   (\lambda \dot{\bf z}_{1{\rm T}} \dot{z}_{j0} ^{\prime}
   + (1-\lambda) \dot{\bf z}_{j {\rm T}} ^{\prime}
   \dot{z}_{j0} )^2 \Big \}^{1\over 2} \, ,
\label{eq:nstrai}
\end{eqnarray}
where we have used $z_j ^{\prime}$ for $z_j (\tau_j ^{\prime})$ and $\epsilon (t)$ 
denotes the sign function. Eq. (\ref{eq:nstrai}) corresponds to span the 
surface by a straight line joining two points with the same time 
coordinate on the quark and the antiquark world lines respectively. The 
sign factors and the second term are necessary to reconstruct the 
surface as the algebraic sum of various pieces when the world-lines go 
backward in time.

     The ``straight line equal time approximation'' consists 
in assuming (\ref{eq:nstrai}) even if $\Gamma_{q \bar q}$ does not stay 
on a plane. Notice that (\ref{eq:nstrai}) gives always 
$S_{\rm min}$ correctly up to the order $(\dot {\bf z}/ \dot z_0)^2$ in 
a semi-relativistic expansion. Furthermore (\ref{eq:nstrai}) 
is exact for particular geometries. This is the case e. g. for 
$\gamma_1$ and $\gamma_2$ making a regular double helix (with axis 
parallel to the time axis) corresponding to a pure rotational motion of the 
quark and the antiquark around a fixed point. Notice also that the approximation 
depends in general on the reference frame. Keeping in mind the helix 
example, however, we shall assume its validity in the center of mass 
frame.

   In a similar way, in the case of $\Gamma _q$ we can set for $x^0=y^0$ 
we can set
\begin{eqnarray}
S_{\rm min}&=&  -\int_0^{s}
 d\tau \int_0^{\tau} d\tau ^{\prime} \delta(z_0-z _0 ^ \prime)
  \vert {\bf z} -{\bf z} ^{\prime}\vert
\epsilon (\dot{z}_0)
\epsilon (\dot{z}_0  ^\prime)
\int_0^1 d\lambda
 \Big \{ {\dot{z}_0^{ 2}} {\dot{z}_{0}^{\prime 2}}-  \nonumber\\
 & & \qquad- (\lambda \dot{\bf z}_{\rm T} \dot{z}_{0} ^{\prime}
 + (1-\lambda) \dot{\bf z}_{\rm T} ^{\prime}
 \dot{z}_{0} )^2 \Big \}^{1\over 2} \, .
\label{eq:rstrai}
\end{eqnarray}

    It is clear that, to include consistently the cases $x_1^0 
\not = x_2^0$, $y_1^0 \not = y_2^0$, $x^0 \not = y^0$, the line integrals in 
(\ref{eq:nstrai}) and (\ref{eq:nstrai}) should be extended in an obvious 
way to the Schwinger strings. However, we find convenient to consider two 
new functions that are obtained replacing (\ref{eq:pertb}) and 
(\ref{eq:nstrai}) in (\ref{eq:hqq}) and (\ref{eq:pertc}) and 
(\ref{eq:rstrai}) in (\ref{eq:hq}) so as they stand. We obtain in this 
way the following equations
\begin{eqnarray}
& & H(x_1,x_2;y_1,y_2) = ({1 \over 2})^2 \int_0^{\infty} d s_1
  \int_0^{\infty} d s_2
  \int_{y_1}^{x_1} {\cal D} z_1\int_{y_2}^{x_2} {\cal D} z_2 \nonumber \\
& &  \exp \Big \{  -{i \over 2} 
  \sum_{j=1}^2  \int_0^{s_j} d\tau_j (m_j^2 +\dot{z}_j^2) \Big \} 
  {\cal S}_0^{s_1} {\cal S}_0^{s_2}
 \exp \Big \{ i \sum_{j=1}^2 \int_0^{s_j} d \tau_j  \int_0^{\tau_j} 
  d \tau _j ^\prime 
    E( z_j -z_j^\prime ; \dot z _j , \dot z _j^\prime) \nonumber\\
& & \qquad \qquad \qquad \qquad  \qquad \qquad -i \int_0^{s_1} 
   d \tau_1  \int_0^{s_2} d \tau _2 
   E( z_1 -z_2 ; \dot z _1 , \dot z _2) \Big \}
\label{eq:fsqq}
\end{eqnarray}
and
\begin{eqnarray}
H(x-y) &=& {1 \over 2} \int_0^{\infty} d s
 \int_y^x {\cal D} z
 \exp \big \{  -{i \over 2} 
 \int_0^{s} d\tau (m^2 +\dot{z}^2) \big \} \nonumber \\
& &  \qquad \qquad \qquad \qquad  {\cal S}_0^s
\exp \big \{i \int_0^s \int_0^\tau 
E(z - z^\prime; \dot z , \dot z ^\prime) \big \} \ ,
\label{eq:fsq}
\end{eqnarray}
where we have set
\begin{equation}
E(\zeta;p,p^\prime) = E_{\rm pert} (\zeta;p,p^\prime) + 
      E_{\rm conf} (\zeta;p,p^\prime) 
\label{eq:eea}
\end{equation}
\noindent
with 
\begin{equation}
\left \{
\begin{array}{ll}
E_{\rm pert} & = 4 \pi {4 \over 3} \alpha_{\rm s} D_{\mu \nu}(\zeta)
   p^\mu p^{\prime \nu}  \nonumber \\
E_{\rm conf} & = \delta (\zeta _0) \vert {\bf \zeta} \vert \epsilon (p_0)
   \epsilon (p_0^{\prime})
   \int_0^1 d \lambda \{ p_0^2 p_0^{\prime 2} - [\lambda p_0^{\prime}
   {\bf p}_{\rm T}  
   + (1- \lambda) p_0 {\bf p}_{\rm T} ^{\prime} ]^2 \}^{1 \over 2} 
\label{eq:eeb}
\end{array} \right.
\end{equation}

    Obviously for arbitrary arguments the quantities $H(x_1, x_2;y_1, y_2)$ 
and $H(x-y)$ as defined by (\ref{eq:fsqq})-(\ref{eq:eeb}) can differ 
very significantly from the original $H^{\rm gi}(x_1, x_2;y_1, y_2)$ and
$H^{\rm gi}(x-y)$.  However, as we mentioned, the two couples coincide
in the limits $x_2 \to x_1$, $y_2 \to y_1$, and $y \to x$ and they are completely 
equivalent for what concerns bound state problems and condensate determination.


\section{Bethe-Salpeter and Dyson-Schwinger equations}

   From eq.s (\ref{eq:fsqq}) and (\ref{eq:fsq}), by various manipulation
and using an appropriate iterative procedure, a Bethe-Salpeter equation
for the function $H(x_1,x_2;y_1,y_2)$ and a Dyson-Schwinger equation for
$H(x-y)$ can be derived in the form \cite{prosp96}
\begin{eqnarray}
H(x_1,x_2;y_1,y_2) &=& H_1(x_1-y_1) \, H_2(x_2-y_2) - \nonumber \\
    & & - i \int d^4 \xi_1 d^4 \xi_2 d^4 \eta_1 d^4 \eta_2 
    H_1(x_1-\xi_1) \, H_2(x_2-\xi_2) \nonumber \\
    & & \qquad \qquad \times I_{ab}(\xi_1,\xi_2;\eta_1,\eta_2)\, 
    \sigma_1^a \, 
    \sigma_2^b \, H(\eta_1,\eta_2;y_1,y_2) \, ,
\label{eq:bseq}
\end{eqnarray}
%
\begin{eqnarray}
H(x-y) &=& H_0(x-y) + i \int d^4 \xi d^4 \eta d^4 \xi^\prime 
       d^4 \eta^\prime H_0(x-\xi) \nonumber \\
       & & \times I_{ab}(\xi,\xi^\prime;\eta,\eta^\prime)
       \sigma^a H(\eta - \eta^\prime) \sigma^b H(\xi^\prime - y) \ ,
\label{eq:sdconf}
\end{eqnarray}
\noindent
where we have set $a, \, b = 0, \, \mu\nu$, with $\sigma^0=1$, and $H_1$ and 
$H_2$ denote the quark and the antiquark H-propagators respectively.

   If we pass to the momentum representation, the corresponding homogeneous
BS-equation becomes in a $4 \times 4$ matrix representation 
\begin{eqnarray}
\Phi_P (k) &=& -i \int {d^4u \over (2 \pi)^4} 
   \hat I_{ab} \big (k-u, {1 \over 2}P
   +{k+u \over 2}, 
   {1 \over 2}P-{k+u \over 2} \big )\nonumber \\
    & & \qquad \qquad \qquad 
   \hat H_1   \big ({1 \over 2} P  + k \big ) 
      \sigma^a  \Phi_P (u) \sigma^b
   \hat H_2 \big (-{1 \over 2} P + k \big ) \, ,
\label{eq:bshoma}
\end{eqnarray}
\noindent
where $\Phi_P (k)$ denotes an appropriate wave function and the center of 
mass frame has to be understood; i.e. $P=(m_B, {\bf 0})$.

     Similarly, in terms of the irreducible self-energy, defined by
$\hat H(k) =\hat H_0(k) + i\hat H_0(k)\hat \Gamma (k) \hat H(k) \,$,
the DS-equation can be written also
\begin{equation}
\hat \Gamma(k) =  \int {d^4 l \over (2 \pi)^4}  \,
\hat I_{ab} \Big ( k-l;{k+l \over 2},{k+l \over 2} \Big ) 
\sigma^a \hat H(l) \, \sigma^b \ .
\label{eq:sdeq}
\end{equation}

   Notice that in principle (\ref{eq:bseq}) and  (\ref{eq:sdconf})
or  (\ref{eq:bshoma}) and  (\ref{eq:sdeq}) are exact equations. However 
the kernels $I_{ab}$ are generated in the form of an expansion in 
$\alpha_{\rm s}$ and $\sigma$. At the lowest order in both such 
constants, we have explicitly
\begin{eqnarray}
& & \hat I_{0;0} (Q; p, p^\prime)  =  4 \int d^4 \zeta \, e^{iQ \zeta} 
   E(\zeta ;p,p^\prime) = 
   16 \pi {4 \over 3} \alpha_{\rm s} p^\alpha p^{\prime \beta}  
  \hat D_{\alpha \beta} (Q)  + \nonumber \\ 
& &  \qquad \qquad + 4 \sigma  \int d^3 {\bf \zeta} e^{-i{\bf Q}
   \cdot {\bf \zeta}} 
    \vert {\bf \zeta} \vert \epsilon (p_0) \epsilon ( p_0^\prime )
   \int_0^1 d \lambda \{ p_0^2 p_0^{\prime 2} -
   [\lambda p_0^\prime {\bf p}_{\rm T} +
   (1-\lambda) p_0 {\bf p}_{\rm T}^\prime ]^2 \} ^{1 \over 2} \nonumber \\
& & \hat I_{\mu \nu ; 0}(Q;p,p^\prime) = 4\pi i {4 \over 3} \alpha_{\rm s} 
   (\delta_\mu^\alpha Q_\nu - \delta_\nu^\alpha Q_\mu) p_\beta^\prime
   \hat D_{\alpha \beta}(Q)  - \nonumber \\
& & \qquad \qquad \qquad  - \sigma  \int d^3 {\bf \zeta} \, e^{-i {\bf Q} \cdot {\bf \zeta}} 
   \epsilon (p_0)
   {\zeta_\mu p_\nu -\zeta_\nu p_\mu \over 
   \vert {\bf \zeta} \vert \sqrt{p_0^2-{\bf p}_{\rm T}^2}} 
   p_0^\prime  \nonumber \\
& & \hat I_{0; \rho \sigma}(Q;p,p^\prime) = 
   -4 \pi i{4 \over 3} \alpha_{\rm s} 
   p^\alpha (\delta_\rho^\beta Q_\sigma - \delta_\sigma^\beta Q_\rho)
   \hat D_{\alpha \beta}(Q) + \nonumber \\
& & \qquad \qquad  \qquad  + \sigma  \int d^3 {\bf \zeta} \, e^{-i{\bf Q} 
  \cdot {\bf \zeta}} p_0 
  {\zeta_\rho p_\sigma^\prime - \zeta_\sigma p_\rho^\prime \over 
  \vert {\bf \zeta} \vert \sqrt{p_0^{\prime 2}
   -{\bf p}_{\rm T}^{\prime 2}} } 
  \epsilon (p_0^\prime)  \nonumber \\
& & \hat I_{\mu \nu ; \rho \sigma}(Q;p,p^\prime) =  
   \pi {4\over 3} \alpha_{\rm s}
  (\delta_\mu^\alpha Q_\nu - \delta_\nu^\alpha Q_\mu) 
  (\delta_\rho^\alpha Q_\sigma - \delta_\sigma^\alpha Q_\rho)
  \hat D_{\alpha \beta}(Q) 
\label{eq:imom}
\end{eqnarray}
\noindent
where in the second and in the third equation $\zeta_0 = 0$ has to be 
understood. 

	Setting            
\begin{equation}
i \hat H^{-1}(k) = \sum_{r=0}^3 \omega_r(k) h_r(k) \, ,
\label{eq:hrecb}
\end{equation}
\noindent
with $\omega_0 = 1, \ \ \omega_1 = \gamma^0 , \ \
\omega_2 = - {\bf \gamma} \cdot \hat{\bf k}, \ \
\omega_3 = \gamma^0 {\bf \gamma} \cdot \hat {\bf k},\ \ \hat {\bf k} = 
{1 \over |{\bf k}|} {\bf k} $ and $h_0(k), 
\dots h_3(k)$ functions of $k_0$ and $|{\bf k}|$. Eq.(\ref{eq:sdeq})
can also be written
\begin{equation}
h_r (k) = \delta_{r0} (k^2-m^2) - i \sum_{s=0}^3 \int {d^4 l \over (2 \pi)^4} 
    {R_{rs}(k,l) \, h_s(l) \over h_0^2(l) - h_1^2(l) + h_2^2(l) -h_3^2(l)} \, ,
\label{eq:sdexpl}
\end{equation}
\noindent
%
\begin{equation}
R_{rs}(k,l) = \mp {1 \over 4} \hat I_{ab} (k-l;{k+l \over 2}, {k+l \over 2})
      {\rm Tr} [ \omega_r^+(k) \sigma^a  \omega_s(k) \sigma^b ] \, ,
\label{eq:kernexpl}
\end{equation}
\noindent
where the sign $-$ applies to the $s=0$ case, the sign $+$ to all the other 
cases. Notice that actually only $ R_{00}, \  R_{11},
\ R_{12}, \ R_{21}, \ R_{22}, \  R_{33} $ are different from zero.

  In connection with the problem of the light pseudo-scalar mesons, let us now
consider eq.s (\ref{eq:bshoma}) and (\ref{eq:sdexpl}) in the chiral limit $m_1=m_2=0$.
As it is apparent from (\ref{eq:qeqg}) chiral symmetry is broken in such limit, if $h_1$
and $h_2$ do not both vanish. On the other side, if this is the case and if
in addition $h_3 \to 0$,
it can be checked that eq. (\ref{eq:bshoma}) is solved for $P=0$ by
$\Phi_0(k)=\hat H_1(k)[\gamma^0 h_1(k) + {\bf \gamma} \ cdot \hat{\bf k}
\, h_2(k)]\gamma^5 \hat H_2(k) $, and a zero mass pseudo-scalar bound state exists,
consistently with the Goldstone theorem. Notice that such a result is
strictly related to the occurrence of the same kernel in the two 
equations  and this in turn is due to the inclusion of 
the self-energy terms in (\ref{eq:nstrai}), i. e. to the correct account 
of world lines that go backwards in time. Obviously the supposed behavior of
the solution is superficially consistent with the form of
(\ref{eq:sdexpl}); presently, however, we are not able to produce any proof.


\section{Three-dimensional reduction and spectrum}

      By replacing $\hat H_1(k)$ and $\hat H_2(k)$ in the BS equation 
by the corresponding free propagators ${i \over k^2-m_j^2}$ and performing 
an instantaneous approximation on the kernels, one can obtain a three
dimensional reduction of the original equation in the form of the eigenvalue
equation for the relativistic Hamiltonian \cite{prosp96} 
\begin{equation}
  H = \sqrt{ m_{1}^{2} + {\bf k}^{2} } +
  \sqrt{ m_{2}^{2} + {\bf k}^{2} } + V,
\label{eq1a}
\end{equation}
with
\begin{eqnarray}
&&  \big< {\bf k} | V | {\bf k'} \big> =
  \frac{1}{2 \sqrt{ w_{1} w_{2}
  w_{1}^{\prime} w_{2}^{\prime} } }
  \bigg\{ \frac{4}{3} \frac{ \alpha_{ \rm{s} } }{\pi^{2}}
  \left[ - \frac{1}{ {\bf Q}^{2} } \left( q_{10}
  q_{20} + {\bf q}^{2} -
  \frac{({\bf Q} \cdot {\bf q})^{2}}{ {\bf Q}^{2} }
  \right) + \right.
\nonumber  \\
&&  + \frac{i}{2 {\bf Q}^{2} } {\bf k'} \times {\bf k} \cdot (
  \mbox{\boldmath $ \sigma $}_{1}
  +
  \mbox{\boldmath $ \sigma $}_{2}
  ) + \frac{1}{2 {\bf Q}^{2}} \bigg[ q_{20} (
  \mbox{\boldmath $ \alpha $}_{1}
  \cdot {\bf Q}) - q_{10} (
  \mbox{\boldmath $ \alpha $}_{2}
  \cdot {\bf Q} ) \bigg] +
\nonumber   \\
&&  +  \frac{1}{6} \left.
  \mbox{\boldmath $ \sigma $}_{1}
  \cdot
  \mbox{\boldmath $ \sigma $}_{2}
  + \frac{1}{4}
  \left( \frac{1}{3}
  \mbox{\boldmath $ \sigma $}_{1}
  \cdot
  \mbox{\boldmath $ \sigma $}_{2} -
  \frac{( {\bf Q} \cdot
  \mbox{\boldmath $ \sigma $}_{1}
  ) ( {\bf Q} \cdot
  \mbox{\boldmath $ \sigma $}_{2}
  )}{ {\bf Q}^{2} }
  \right) +
  \frac{1}{4 {\bf Q}^{2} } (
  \mbox{\boldmath $ \alpha $}_{1}
  \cdot {\bf Q}) (
  \mbox{\boldmath $ \alpha $}_{2}
  \cdot {\bf Q} ) \right]  \nonumber \\
&&  +  \frac{1}{( 2 \pi )^{3}}
  \int \! d^{3} {\bf r} \, e^{i {\bf Q} \cdot {\bf r} }
  J^{ \rm{inst} }({\bf r},{\bf q},q_{10},q_{20}) \bigg\} \,,
\label{eq:due}
\end{eqnarray}
%
\begin{eqnarray}
&&  J^{ \rm{inst} }({\bf r},{\bf q},q_{10},q_{20}) =
  \frac{ \sigma r }{ q_{10} + q_{20} }
  \bigg[ q_{20}^{2} \sqrt{ q_{10}^{2} - {\bf q}_{ \rm{T} }^{2} } +
  q_{10}^{2} \sqrt{ q_{20}^{2} - {\bf q}_{ \rm{T} }^{2} } +
\nonumber \\
&&  + \frac{ q_{10}^{2} q_{20}^{2} }{ | {\bf q}_{ \rm{T} } | }
  \bigg( \arcsin \frac{ | {\bf q}_{ \rm{T} } | }{q_{10}} +
  \arcsin \frac{ | {\bf q}_{ \rm{T} } | }{q_{20}} \bigg) \bigg]
  - \frac{ \sigma }{ r } \bigg[
  \frac{ q_{20} }{ \sqrt{ q_{10}^{2} - {\bf q}_{ \rm{T} }^{2} } }
  \bigg( {\bf r} \times {\bf q} \cdot
  \mbox{\boldmath $ \sigma $}_{1}
  + i q_{10} ( {\bf r} \cdot
  \mbox{\boldmath $ \alpha $}_{1}
  ) \bigg) 
\nonumber \\
&&  + \frac{ q_{10} }{ \sqrt{ q_{20}^{2} - {\bf q}_{ \rm{T} }^{2} } }
  \bigg( {\bf r} \times {\bf q} \cdot
  \mbox{\boldmath $ \sigma $}_{2}
  - i q_{20} ( {\bf r} \cdot
  \mbox{\boldmath $ \alpha $}_{2}
  ) \bigg) \bigg] .
\label{eq:tre}
\end{eqnarray}
In eq.s (\ref{eq1a}-\ref{eq:tre})
$ {\bf k'} $ and $ {\bf k} $ denote
the final and the initial center of mass
momentum of the quark;
$ w_{j} = \sqrt{ m_{j}^{2} + {\bf k}^{2} }$,
$ w_{j}^{\prime} = \sqrt{ m_{j}^{2} + {\bf k}^{\prime 2} }$,
$ {\bf q} = \frac{ {\bf k} + {\bf k'} }{2}$,
$ {\bf Q} = {\bf k'} - {\bf k}$,
$ q_{j0} = \frac{ w_{j} + w_{j}^{\prime} }{2}$; 
$ q_{ \rm{T} }^{h} = ( \delta^{h k} - \hat{r}^{h} \hat{r}^{k} ) q^{k}$
is the transverse momentum,
$ \alpha^{k} $ are the usual Dirac matrices $ \gamma^{0}
\gamma^{k} $, and $ \sigma^{k} =  1/2 \, \varepsilon^{knm} \sigma^{nm}$
the $ 4 \times 4 $ spin matrices.

   In spite of its complication the above expression has various significant
limit cases that corresponds to models successfully used in different areas.
In the static limit it gives the local potential 
\begin{equation}
V = - {4 \over 3} {\alpha_{\rm s} \over r } + \sigma r \, .
    \label{eq:static}
\end{equation}
%
%
In the heavy
masses limit, by an ${1 \over m}$ expansion and an
appropriate Foldy-Wouthuysen transformation, it reduces to the potential
discussed in ref. \cite{prspcomo1}, If the spin dependent terms are
neglected, $V$ becomes identical (apart from a question of ordering) to the first order
expansion of the potential
corresponding to the relativistic flux tube model \cite{prspcomo1,tubolsson}.

     We have attempt to apply the  actual $V$ as given by (\ref{eq:due}) 
and (\ref{eq:tre}) to the determination of the spectrum. In the preliminary
calculation we have performed up to now
the spin orbit terms 
have been omitted, due to their complication; however, the hyperfine 
terms have been included.

\begin{figure}
\begin{centering}
    \leavevmode
    \setlength{\unitlength}{1.0mm}
    \begin{picture}(140,70)
      \put(25,0){\mbox{\epsfig{file=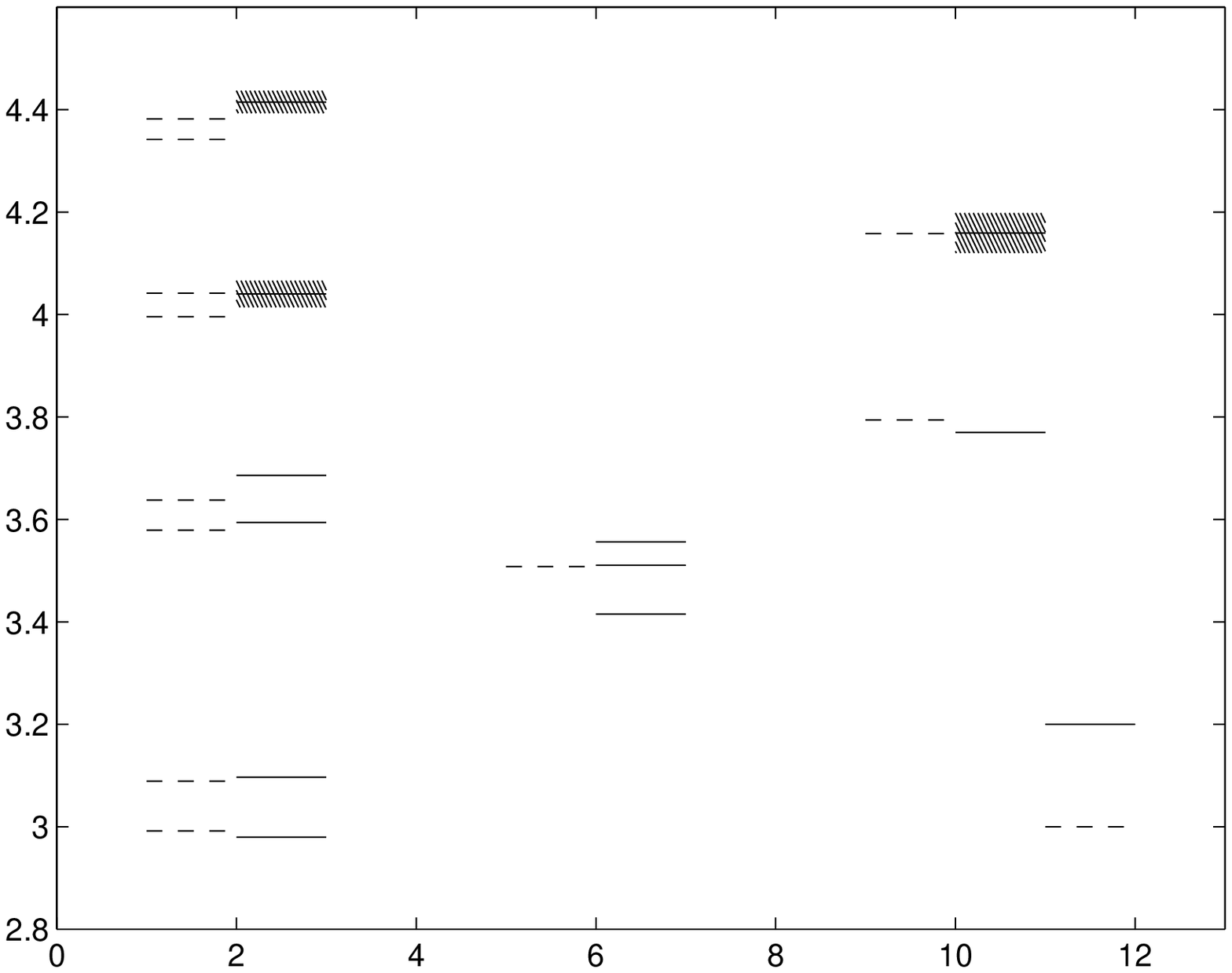,height=7cm}}}
      \put(16,65){GeV}
      \put(16,55){M}
      \put(70,61){ \Large{$ c \bar{c} $ } }
      \put(88,10){theor.}
      \put(88,17){exper.}
      \put(48,8){ $ 1 \, {^{1} {\rm S}_{0}} $ }
      \put(48,13){ $ 1 \, {^{3} {\rm S}_{1}} $ }
      \put(48,30){ $ 2 \, {^{1} {\rm S}_{0}} $ }
      \put(48,35){ $ 2 \, {^{3} {\rm S}_{1}} $ }
      \put(48,47){ $ 3 \, {^{3} {\rm S}_{1}} $ }
      \put(48,61){ $ 4 \, {^{3} {\rm S}_{1}} $ }
      \put(74,23){ $ 1 \, {^{3} {\rm P}_{0}} $ }
      \put(74,27){ $ 1 \, {^{3} {\rm P}_{1}} $ }
      \put(74,31){ $ 1 \, {^{3} {\rm P}_{2}} $ }
      \put(100,38){ $ 1 \, {^{3} {\rm D}_{1}} $ }
      \put(100,52){ $ 2 \, {^{3} {\rm D}_{1}} $ }
    \end{picture}
\caption{ Charmonium spectrum. }
\label{fig4}
\end{centering}
\end{figure}
%

%
\begin{figure}
\begin{centering}
    \leavevmode
    \setlength{\unitlength}{1.0mm}
    \begin{picture}(140,70)
      \put(25,0){\mbox{\epsfig{file=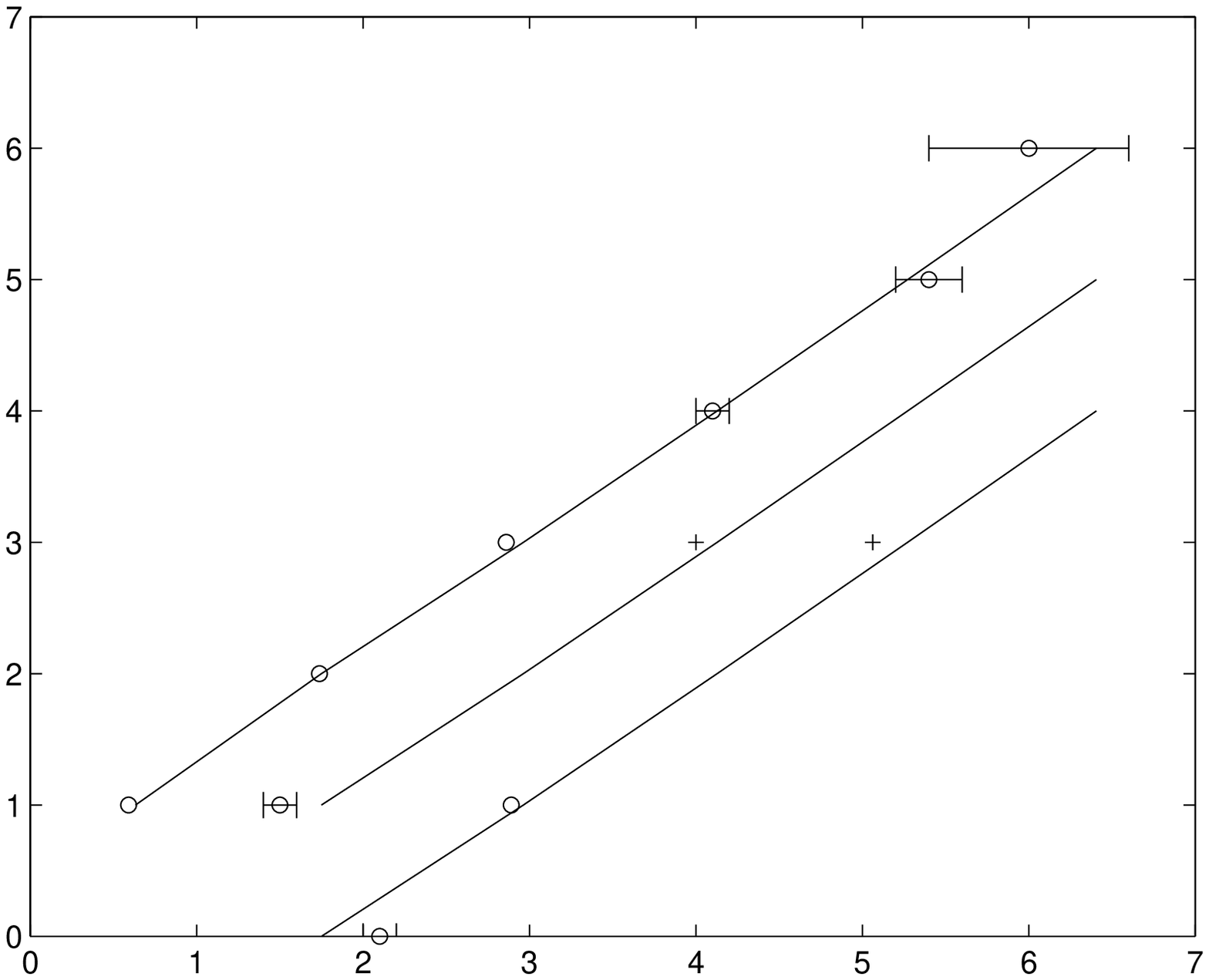,height=7cm}}}
      \put(16,60){ $ J $ }
      \put(110,59){ $ J = L + 1 $ }
      \put(110,49){ $ J = L $ }
      \put(110,39){ $ J = L - 1 $ }
      \put(45,60){ $ \alpha^{ \prime } = 0.878 $ }
      \put(111,0){ $ M^{2} $ }
      \put(20,14){ $ \rho ( 770 ) $ }
      \put(28,22){ $ a_{2} ( 1320 ) $ }
      \put(32,7){ $ a_{1} ( 1260 ) $ }
      \put(40,-3){ $ a_{0} ( 1450 ) $ }
      \put(41,30){ $ \rho_{3} ( 1690 ) $ }
      \put(59,9){ $ \rho ( 1700 ) $ }
      \put(56,40){ $ a_{4} ( 2040 ) $ }
      \put(61,26){ $ X ( 2000 ) $ }
      \put(70,50){ $ \rho_{5} ( 2350 ) $ }
      \put(84,26){ $ \rho_{3} ( 2250 ) $ }
      \put(90,62){ $ a_{6} ( 2450 ) $ }
    \end{picture}
\caption{Ground triplet $ u \bar{u} $ Regge trajectories.
Theoretical results (full line) compared with experimental
data (circlet). Cross denote less established masses.}
\label{fig1}
\end{centering}
\end{figure}

\begin{table}
\caption{Theoretical results for
$ u \bar{c} $, $ u \bar{b} $, $ s \bar{c} $, $ s \bar{b} $ systems
(MeV). Experimental data are enclosed in brackets.}
\begin{tabular}{ccccc}
\hline
 State & $ u \bar{c} $ & $ u \bar{b} $ &
 $ s \bar{c} $ & $ s \bar{b} $ \\
\hline
 1S &
 1973 ($ 1973 \pm 1 $) &
 5326 ($ 5313 \pm 2 $) &
 2080 ($ 2076.4 \pm 0.5 $) &
 5418 ($ 5404.6 \pm 2.5) $
\\
 2S &
 2600 $ ( 2623 \pm ? )^{\rm a} $ &
 5906 $ ( 5897 \pm ? )^{\rm a} $ &
 2713 &
 6004
\\
 1P &
 2442 $ ( 2438 \pm ? )^{\rm b} $ &
 5777 $ ( 5825 \pm 14 )^{\rm c} $ &
 2528 ($ 2535.35 \pm 0.34 $)  &
 5848 ($ 5853 \pm 15 $)
\\
\hline
\end{tabular}
\label{tablegn}
$ ^{\rm a} $Obtained from preliminary {\sl Delphi} data
$ m(D^{\ast \prime}) = 2637 \pm 8 $ MeV,
$ m(B^{\ast \prime}) = 5906 \pm 14 $ MeV \cite{pullia}
subtracting 1/4 theoretical hyperfine splitting reported in
table \ref{tableg8}.
\\
$ ^{\rm b} $Estimated from
$ m(D_{2}^{\ast}) = 2459 \pm 4 $ MeV,
$ m(D_{1}) = 2427 \pm 5 $ MeV.
\\
$ ^{\rm c} $From preliminary {\sl Delphi} data \cite{pullia}.
\end{table}
\begin{table}
\centering
\caption{Theoretical results for 
$ q \bar{q} $ hyperfine splitting (MeV).
Experimental data are enclosed in brackets.}
\begin{tabular}{ccccccc}
\hline
 State & $ u \bar{c} $ & $ u \bar{b} $ & $ c \bar{c} $ &
 $ b \bar{b} $ & $ s \bar{c} $ & $ s \bar{b} $ \\
\hline
 1S &
 111 ($ 141 \pm 1 $) &
 59 ($ 46 \pm 3 $) &
 97 ($ 117 \pm 2 $) &
 102 &
 108 (144) &
 60 ($ 47 \pm 4) $    \\
 2S &
 59 &
 38 &
 59 ($ 92 \pm 5 $) &
 42 &
 62 &
 40 \\
\hline
\end{tabular}
\label{tableg8}
\end{table}

    The numerical procedure we have followed consists in solving first the 
eigenvalue equation for the static potential (\ref{eq:static}) by 
the Rayleigh-Ritz method  and then in evaluating the quantities $\langle H 
\rangle$ for the eigenfunctions obtained in the first step.
We have adopted the following parameters:
$ \alpha_{\rm s} = 0.363 $,
$ \sigma = 0.175 $ GeV$^{2} $,
$ m_{c} = 1.405 $ GeV,
$ m_{b} = 4.81 $ GeV,
$ m_{s} = 200 $ MeV,
$ m_{u} = 10 $ MeV.
The first four values have to be compared with those
obtained from heavy quarkonium fits (e.g. \cite{tstrlt,tubolsson}) and 
apart from the possibility of a small rearrangement are completely 
determined by these. The light quark masses have been fixed a priori on 
typical current values as reported by the Particle Data Group.

    By such parameters one succeeds to reproduce reasonably well the 
not only the bottonium and the charmonium spectrum, but also the Regge 
trajectories (with correct slope and intercepts) for the 
ground triplet states of the $u \bar u$, $u \bar s$, $s \bar s$  systems, 
and the known spin averaged states for the light-heavy systems. Notice that 
no ad hoc constant has been added to the potential and that in particular 
the heavy-light sector is completely parameter free.

    Our results have been reported in full in ref \cite{mio}, where even
some details on the the numerical difficulties
are explained. Here as an example for the heavy-heavy systems we 
report in fig. \ref{fig4} the charmonium spectrum and
for the light-light systems we report in fig.
\ref{fig1} the $\rho$ Regge trajectory.
The results for heavy-light systems
are reported in table \ref{tablegn} and compared with the
experimental spin averaged masses by using the theoretical
splitting of table \ref{tableg8}
where the singlet states are not available.

%

%
\end{document}